\documentclass[twocolumn,nofootinbib,amsmath,amssymb,a4paper]{revtex4}

\usepackage{graphicx}
\usepackage{dcolumn}
\usepackage{bm}

\RequirePackage{xspace}

\usepackage{relsize}
\def\babar{\mbox{\slshape B\kern-0.1em{\smaller A}\kern-0.1em
    B\kern-0.1em{\smaller A\kern-0.2em R}}}

\def\epem       {\ensuremath{e^+e^-}\xspace}

\def\q     {\ensuremath{q}\xspace}

\def\qqbar {\ensuremath{q\overline q}\xspace}
\def\u     {\ensuremath{u}\xspace}

\def\d     {\ensuremath{d}\xspace}

\def\s     {\ensuremath{s}\xspace}

\def\c     {\ensuremath{c}\xspace}
\def\cbar  {\ensuremath{\overline c}\xspace}
\def\ccbar {\ensuremath{c\overline c}\xspace}
\def\b     {\ensuremath{b}\xspace}

\def\t     {\ensuremath{t}\xspace}

\def\piz   {\ensuremath{\pi^0}\xspace}

\def\pim   {\ensuremath{\pi^-}\xspace}

\def\Kbar  {\kern 0.2em\overline{\kern -0.2em K}{}\xspace}

\def\Kz    {\ensuremath{K^0}\xspace}
\def\Kzb   {\ensuremath{\Kbar^0}\xspace}
\def\KzKzb {\ensuremath{\Kz \kern -0.16em \Kzb}\xspace}
\def\Kp    {\ensuremath{K^+}\xspace}
\def\Km    {\ensuremath{K^-}\xspace}

\def\KpKm  {\ensuremath{\Kp \kern -0.16em \Km}\xspace}
\def\KS    {\ensuremath{K^0_{\scriptscriptstyle S}}\xspace} 
\def\KL    {\ensuremath{K^0_{\scriptscriptstyle L}}\xspace} 
\def\Kstarz  {\ensuremath{K^{*0}}\xspace}

\def\Kstar   {\ensuremath{K^*}\xspace}

\def\Dbar    {\kern 0.2em\overline{\kern -0.2em D}{}\xspace}

\def\Dz      {\ensuremath{D^0}\xspace}
\def\Dzb     {\ensuremath{\Dbar^0}\xspace}
\def\DzDzb   {\ensuremath{\Dz {\kern -0.16em \Dzb}}\xspace}
\def\Dp      {\ensuremath{D^+}\xspace}
\def\Dm      {\ensuremath{D^-}\xspace}

\def\DpDm    {\ensuremath{\Dp {\kern -0.16em \Dm}}\xspace}

\def\Dstarz  {\ensuremath{D^{*0}}\xspace}

\def\Dstarp  {\ensuremath{D^{*+}}\xspace}
\def\Dstarm  {\ensuremath{D^{*-}}\xspace}

\def\B       {\ensuremath{B}\xspace}
\def\Bbar    {\kern 0.18em\overline{\kern -0.18em B}{}\xspace}

\def\Bz      {\ensuremath{B^0}\xspace}
\def\Bzb     {\ensuremath{\Bbar^0}\xspace}
\def\BzBzb   {\ensuremath{\Bz {\kern -0.16em \Bzb}}\xspace}
\def\Bu      {\ensuremath{B^+}\xspace}
\def\Bub     {\ensuremath{B^-}\xspace}

\def\Bpm     {\ensuremath{B^\pm}\xspace}

\def\BpBm    {\ensuremath{\Bu {\kern -0.16em \Bub}}\xspace}

\def\BorBbar    {\kern 0.18em\optbar{\kern -0.18em B}{}\xspace}
\def\DorDbar    {\kern 0.18em\optbar{\kern -0.18em D}{}\xspace}
\def\KorKbar    {\kern 0.18em\optbar{\kern -0.18em K}{}\xspace}

\def\jpsi     {\ensuremath{{J\mskip -3mu/\mskip -2mu\psi\mskip 2mu}}\xspace}
\def\psitwos  {\ensuremath{\psi{(2S)}}\xspace}

\def\etac     {\ensuremath{\eta_c}\xspace}

\def\chicone  {\ensuremath{\chi_{c1}}\xspace}

\mathchardef\Upsilon="7107
\def\Y#1S{\ensuremath{\Upsilon{(#1S)}}\xspace}

\def\FourS {\Y4S}

\mathchardef\Deltares="7101
\mathchardef\Xi="7104
\mathchardef\Lambda="7103
\mathchardef\Sigma="7106
\mathchardef\Omega="710A

\def\Deltabar{\kern 0.25em\overline{\kern -0.25em \Deltares}{}\xspace}
\def\Lbar{\kern 0.2em\overline{\kern -0.2em\Lambda\kern 0.05em}\kern-0.05em{}\xspace}
\def\Sigbar{\kern 0.2em\overline{\kern -0.2em \Sigma}{}\xspace}
\def\Xibar{\kern 0.2em\overline{\kern -0.2em \Xi}{}\xspace}
\def\Obar{\kern 0.2em\overline{\kern -0.2em \Omega}{}\xspace}
\def\Nbar{\kern 0.2em\overline{\kern -0.2em N}{}\xspace}
\def\Xb{\kern 0.2em\overline{\kern -0.2em X}{}\xspace}

\def\bpsikst    {\ensuremath{\Bz \to \jpsi \Kstar}\xspace}

\def\mes        {\mbox{$m_{\rm ES}$}\xspace}

\newcommand{\tev}{\ensuremath{\mathrm{\,Te\kern -0.1em V}}\xspace}
\newcommand{\gev}{\ensuremath{\mathrm{\,Ge\kern -0.1em V}}\xspace}
\newcommand{\mev}{\ensuremath{\mathrm{\,Me\kern -0.1em V}}\xspace}
\newcommand{\kev}{\ensuremath{\mathrm{\,ke\kern -0.1em V}}\xspace}
\newcommand{\ev}{\ensuremath{\mathrm{\,e\kern -0.1em V}}\xspace}
\newcommand{\gevc}{\ensuremath{{\mathrm{\,Ge\kern -0.1em V\!/}c}}\xspace}
\newcommand{\mevc}{\ensuremath{{\mathrm{\,Me\kern -0.1em V\!/}c}}\xspace}
\newcommand{\gevcc}{\ensuremath{{\mathrm{\,Ge\kern -0.1em V\!/}c^2}}\xspace}
\newcommand{\mevcc}{\ensuremath{{\mathrm{\,Me\kern -0.1em V\!/}c^2}}\xspace}

\def\mus  {\ensuremath{\rm \,\mus}\xspace}

\def\ps   {\ensuremath{\rm \,ps}\xspace}

\def\mus        {\ensuremath{\,\mu{\rm s}}\xspace}    
\def\ps         {\ensuremath{{\rm \,ps}}\xspace}  

\def\rad{\ensuremath{\rm \,rad}\xspace}

\def\to                 {\ensuremath{\rightarrow}\xspace}

\def\pep2{PEP-II}

\def\gsim{{~\raise.15em\hbox{$>$}\kern-.85em
          \lower.35em\hbox{$\sim$}~}\xspace}
\def\lsim{{~\raise.15em\hbox{$<$}\kern-.85em
          \lower.35em\hbox{$\sim$}~}\xspace}

\def\eps{\varepsilon\xspace}

\def\CP                {\ensuremath{C\!P}\xspace}
\def\C       {\ensuremath{C}\xspace}

\def\rhobar {\ensuremath{\overline \rho}\xspace}
\def\etabar {\ensuremath{\overline \eta}\xspace}

\def\stwob{\ensuremath{\sin\! 2 \beta   }\xspace}

\def\mistag{\ensuremath{w}\xspace}

\def\deltaz{\ensuremath{{\rm \Delta}z}\xspace}
\def\deltat{\ensuremath{{\rm \Delta}t}\xspace}
\def\deltamd{\ensuremath{{\rm \Delta}m_d}\xspace}

\xspace

\newcommand{\jprlBase}       {Phys.\ Rev.\ Lett.\xspace}
\newcommand{\jprBase}        {Phys.\ Rev.\xspace}
\newcommand{\jplBase}        {Phys.\ Lett.\xspace}
\newcommand{\nimBaseA}       {Nucl.\ Instrum.\ Methods Phys.\ Res., Sect.\ A\xspace}

\newcommand{\npBase}         {Nucl.\ Phys.\xspace}

\newcommand{\nima}      [1]  {\nimBaseA~{\bf #1}}

\newcommand{\npb}       [1]  {\npBase\ B~{\bf #1}}

\newcommand{\plb}       [1]  {\jplBase\ B~{\bf #1}}

\newcommand{\jprl}      [1]  {\jprlBase\ {\bf #1}}
\newcommand{\jprd}      [1]  {\jprBase\ D~{\bf #1}}

\def\jetset74   {\mbox{\tt Jetset \hspace{-0.5em}7.\hspace{-0.2em}4}\xspace}

\def\leptontag{{\tt Lepton}}
\def\kaonitag{{\tt Kaon\,I}}
\def\kaoniitag{{\tt Kaon\,II}}
\def\kpitag{{\tt Kaon-Pion}}
\def\piontag{{\tt Pion}}
\def\othertag{{\tt Other}}
\def\nbb{347.5 \times 10^6\, \FourS \to B\Bbar}
\def\nbbddk{208.7 \times 10^6\, \FourS \to B\Bbar}
\def\nbbjpsik{88.0 \times 10^6\, \FourS \to B\Bbar}

\newcommand\Acp {\ensuremath{A_{\CP}}}

\def\to           {\ensuremath{\rightarrow}}
\def\B            {\ensuremath{B}}
\def\Bz           {\ensuremath{B^{0}}}
\def\Bzb          {\ensuremath{\overline{B}^{0}}}
\def\piz          {\ensuremath{\pi^{0}}}

\def\bztojpsipiz  {\ensuremath{\Bz \to \jpsi \piz }}

\def\jpsikstarza  {\ensuremath{\jpsi\Kstarz (\Kstarz \to \KS\piz)}}
\newcommand{\hz}{\ensuremath{h^0}\xspace}
\def\Bztojpsikstar {\ensuremath{\Bz \to \jpsikstarza}}
\def\Bztodstdstks  {\ensuremath{\Bz \to \Dstarp \Dstarm \KS}\xspace}

\def\Bztodsthz     {\ensuremath{\Bz \to \Dstarz \hz}\xspace}
\def\Jc           {\ensuremath{J_{c}}}
\def\Jzero        {\ensuremath{J_{0}}}
\def\Jsone        {\ensuremath{J_{s1}}}
\def\Jstwo        {\ensuremath{J_{s2}}}
\def\effectiveeta{0.504 \pm 0.033}
\def\CP      {\ensuremath{ CP }}

\def\fbar  {\ensuremath{\overline f}\xspace}
\def\C       {\ensuremath{ C }}
\def\S       {\ensuremath{ S }}

\def\mes     { \ensuremath { m_{ES} } }
\def\deltat  { \ensuremath{\Delta t }}
\def\deltaz  { \ensuremath{\Delta z }}

\def\deltamd { \ensuremath{\Delta m_d }}

\def\epem      {\ensuremath{ { e^+ e^- } } }
\def\qqbar     {\ensuremath{ { q \overline{q} } } }

\def\btoccbars{\ensuremath{\b\rightarrow\c\cbar\s}}
\def\ctwob{\ensuremath{\cos\! 2 \beta   }\xspace}
\def\fitstwob{0.710}
\def\statstwob{0.034}
\def\syststwob{0.019}
\def\fitlambda{0.932}
\def\statlambda{0.026}
\def\systlambda{0.017}
\newcommand{\bflav}{\ensuremath{\B_{{\rm flav}}}}

\newcommand{\SLACPubNumber} {12316}

\begin{document}

\title{
\begin{flushleft}
SLAC-PUB-\SLACPubNumber\\
\end{flushleft}
\vskip 5pt
Measurements of $\stwob$ and $\ctwob$ from $\btoccbars$ decays at \babar.}

\author{Katherine A. George}
\email{katherine.george@slac.stanford.edu}
\affiliation{Department of Physics, Queen Mary, University of London, Mile End Road, London, E1 4NS, UK.}

\begin{abstract}
{Recent measurements of \stwob\ and \ctwob\ using \btoccbars\ decays are presented using data collected by the \babar\ experiment at
the PEP-II asymmetric-energy $B$-factory.}
\end{abstract}

\maketitle

\section{Introduction.}
\label{sec:intro}
The Standard Model (SM) of particle physics describes charge conjugation-parity (\CP) violation as a consequence of a 
complex phase in the three-generation Cabibbo-Kobayashi-Maskawa (CKM) quark-mixing matrix~\cite{ref:ckm}. 
In this framework, measurements of \CP\ asymmetries in the proper-time distribution of neutral $B$ decays to 
\CP\ eigenstates containing a charmonium and $K^{0}$ meson provide a direct measurement of $\stwob$~\cite{ref:BCP}. 
The unitarity triangle angle $\beta$ is $\arg \left[\, -V_{\rm cd}^{}V_{\rm cb}^* / V_{\rm td}^{}V_{\rm tb}^*\, \right]$ where
the $V_{ij}$ are CKM matrix elements. \CP\ violation in the $B$-meson system was established by the \babar\ ~\cite{ref:babar-stwob-prl}
and Belle~\cite{ref:belle-stwob-prl} collaborations in 2001.
 
The \babar\ detector~\cite{ref:babar} is located at the SLAC PEP-II $e^+e^-$ asymmetric energy \B -factory~\cite{ref:pep}
where data is collected on or just below the $\FourS$ resonance. A small fraction ($\approx$ 10$\%$) is collected
at approximately 40 MeV below the $\FourS$ resonance, and is used to study background from \epem\to\qqbar ($\q = \u,\d,\s,\c$) continuum events.
The \babar\ experimental program includes the measurement of the angle $\beta$ through the measurement of time-dependent $\CP$-asymmetries, 
$\Acp$ as discussed in Ref.~\cite{ref:babarprd}. $\Acp$ is defined as
\begin{eqnarray}
\Acp(t) & \equiv & \frac{N(\Bzb(t)\to f) - N(\Bz(t)\to f)} {N(\Bzb(t)\to f) + N(\Bz(t)\to f)} \nonumber \\
        &&{} = \S \sin(\deltamd{t}) - \C \cos(\deltamd{t}),
\label{eq:timedependence}
\end{eqnarray}
where $N(\Bzb(t)\to f)$ is the number of \Bzb\ that decay into the $CP$-eigenstate $f$ after a time $t$
and $\deltamd$ is the difference between the \B\ mass eigenstates.
The sinusoidal term describes interference between mixing and decay and the cosine term is the
direct \CP\ asymmetry. 
In Eq.~\ref{eq:timedependence}, $A(\Bzb(t)\to \fbar)$ ($A(\Bz(t)\to\overline{f})$) is the decay amplitude of
$\Bzb$ ($\Bz$) to the final state $\overline{f}$ ($f$).

In this article, the current status of measurements of $\stwob$ and $\ctwob$ from $\btoccbars$ 
decays at \babar\ are discussed. All results are final unless otherwise stated. 
Additional results on \CP\ violation measurements in \B\ to charm decays at
\babar\ can be found in Ref.~\cite{ref:ichepprocs}.

\section{\stwob\ from $\Bz\to$ charmonium + $\Kz$}
\label{sec:stwob}
The determination of $\beta$ from $\btoccbars$ decay modes currently provides the most stringent constraint
on the unitarity triangle.
For these decay modes, the \CP\ violation parameters $\S$ and $\C$ are
$\S_{\btoccbars} = -\eta_{f}\stwob$ and $\C_{\btoccbars}$ = 0, where
$\eta_{f}$ is $-$1 for ($\ccbar$)$\KS$ decays (e.g. $\jpsi\KS$, $\psitwos\KS$, $\chicone\KS$, $\eta_c \KS$
~\cite{ref:charge}) and $\eta_{f}$ is $+$1 for the ($\ccbar$)$\KL$ (e.g. $\jpsi\KL$) state.
We use the value $\eta_f = \effectiveeta$ for the $\jpsi\Kstarz (\Kstarz \to \KS\piz)$ final 
state since it can be both \CP\ even and \CP\ odd due to the presence of even and odd orbital 
angular momentum contributions~\cite{ref:rperp}. These modes have most recently been used to measure 
$\stwob$ using a sample of $\nbb$ decays~\cite{ref:sin2b}. This result is preliminary.

In addition to the \CP\ modes described above, a large sample \bflav\ of \Bz\ decays to the flavor
eigenstates $D^{(*)-}h^+ (h^+=\pi^+,\rho^+$, and $a_1^+)$ and $\jpsi\Kstarz
(\Kstarz\to\Kp\pim)$ is used for calibrating the flavor tagging performance and \deltat\ resolution.
Studies are performed to measure apparent \CP\ violation from unphysical sources using a 
control sample of $B^+$ mesons decaying to the final states $\jpsi K^{(*)+}$, $\psitwos
K^+$, $\chicone \Kp$, and $\eta_c \Kp$. 
The event selection and candidate reconstruction are unchanged from those described in 
Refs.~\cite{ref:babar2004,ref:babarprd,ref:etacks} with the exception of a new 
$\eta_c \KS$ event selection based on the Dalitz structure of the $\etac \to \KS \Kp \pim$ decay.
We calculate the time interval \deltat\ between the two $B$ decays from the measured separation 
\deltaz\ between the decay vertices of $B_{\rm rec}$ and $B_{\rm tag}$ along the collision 
($z$) axis~\cite{ref:babarprd}. The $z$ position of the $B_{\rm rec}$ vertex is determined from
the charged daughter tracks. The $B_{\rm tag}$ decay vertex is determined by fitting tracks 
not belonging to the $B_{\rm rec}$ candidate to a common vertex, employing constraints from the beam spot
location and the $B_{\rm rec}$ momentum~\cite{ref:babarprd}. 
Events are accepted if the calculated \deltat\ uncertainty is less than 2.5\ps
and $\vert \deltat \vert$ is less than  $20\,\ps$. The fraction of events satisfying 
these requirements is 95\%.

At the $\Upsilon(4S)$ resonance, $\Acp$ is extracted from the distribution of the
difference of the proper decay times, $\t \equiv \t_{\CP} - \t_{tag}$, where
$\t_{\CP}$ refers to the decay time of the signal \B\ meson ($\B_{\CP}$) and $\t_{tag}$ refers to the
decay time of the other \B\ meson in the event ($\B_{tag}$). 
Multivariate algorithms are used to identify signatures of $B$ decays that determine 
(``tag'') the flavor of the $B_{\rm tag}$ at decay to be either a \Bz\ or \Bzb candidate. 
These algorithms account for correlations among different sources of flavor information and 
provide an estimate of the mistag probability for each event. 
Each event whose estimated mistag probability is less than 45\% is assigned to one of six 
tagging categories.
The \leptontag\ category contains events with an identified lepton;
the remaining events are divided into the 
\kaonitag, \kaoniitag, \kpitag, \piontag, or \othertag\ categories based on
the estimated mistag probability.  
For each category $i$, the tagging efficiency $\eps_i$ and 
fraction $\mistag_i$ of events having the wrong tag assignment are measured from data. 
The figure of merit for tagging is the effective tagging efficiency
$Q \equiv \sum_i {\eps_i (1-2\mistag_i)^2} = (30.4 \pm 0.3)\,\%$, where the error shown is statistical only.

With the exception of the $\jpsi\KL$ mode, we use the beam-energy substituted mass 
$\mes=\sqrt{{(E^{*}_{\rm beam})^2}-(p^{*}_B)^2}$
to determine the composition of our final sample, where $E^{*}_{\rm beam}$ and $p_B^{*}$ are the 
beam energy and $B$ momentum in the $\epem$ center-of-mass frame.
For the $\jpsi\KL$ mode we use the difference $\Delta E$ between the candidate 
center-of-mass energy and $E^{*}_{\rm beam}$. 
We use events with $\mes > 5.2 \gevcc$ ($\Delta E < 80\mev$ for $\jpsi \KL$) 
in order to determine the properties of the background contributions.
We define a signal region $5.27 < \mes < 5.29 \gevcc$  
($\vert \Delta E \vert < 10\mev$ for $\jpsi \KL$)
that contains {\bf{$ $}} \CP\ candidate events that satisfy the tagging and vertexing 
requirements as listed in Table~\ref{tab:result}.

\begin{figure}[!h]
\begin{center}
\begin{center}
\scalebox{0.38}{\includegraphics{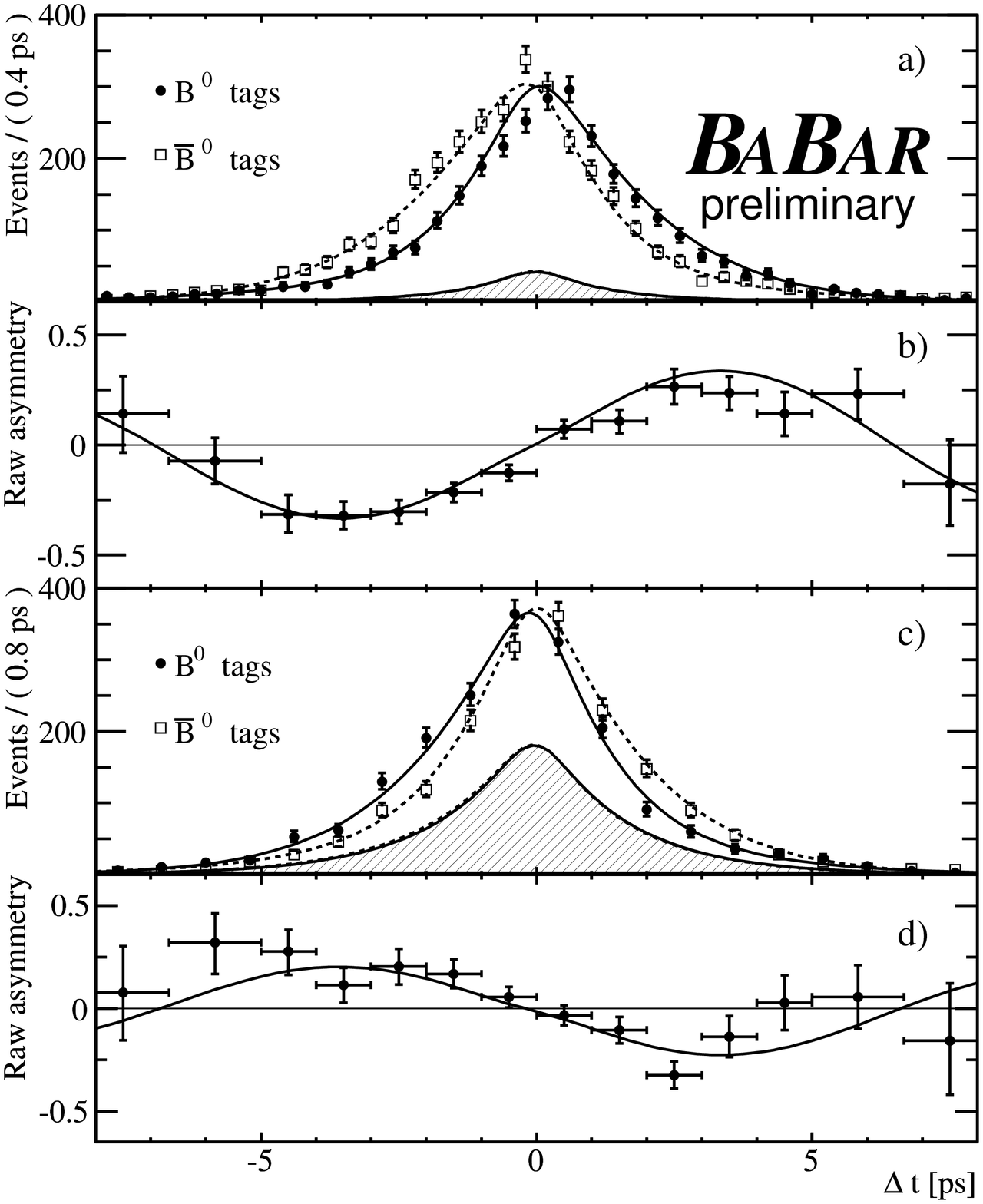}}
\end{center}
\caption{
a) Number of $\eta_f=-1$ candidates ($J/\psi \KS$, $\psi(2S) \KS$, $\chicone \KS$, and $\eta_c \KS$)
in the signal region with a \Bz tag ($N_{\Bz }$) and with a \Bzb tag ($N_{\Bzb}$), and 
b) the raw asymmetry $(N_{\Bz}-N_{\Bzb})/(N_{\Bz}+N_{\Bzb})$, as functions of \deltat.
Figures c) and d) are the corresponding distributions for the $\eta_f=+1$ mode $J/\psi \KL$.  
All distributions exclude \othertag-tagged events. The solid (dashed) curves 
represent the fit projections
in \deltat for \Bz (\Bzb) tags. The shaded regions represent the estimated background contributions.
\label{fig:cpdeltat}}
\end{center}
\end{figure}

\begin{table}[!htb] 
\begin{center}
\caption{
Number of events $N_{\rm tag}$ in the signal region after tagging and vertexing requirements, 
signal purity $P$ including the contribution from peaking background,
and results of fitting for \CP\ asymmetries in the $B_{\CP}$ sample and various subsamples.
In addition, results on the $B_{\rm flav}$ and charged $B$ control samples
test that no artificial \CP\ asymmetry is found where we expect no \CP\ violation ($\stwob=0$).
Errors are statistical only. The  signal region is $5.27 < \mes < 5.29 \gevcc$
 ($\vert \Delta E \vert < 10\mev$ for $\jpsi \KL$).\label{tab:result}}
\vspace{0.5cm}
\begin{tabular}{lrcc}\hline\hline
Sample  & $N_{\rm tag}$ & $P(\%)$ & \multicolumn{1}{c}{$\ \ \ \stwob$}\\ 
Full \CP\ sample                                    & $11496$   & $76$ &  $0.710\pm0.034$   \\\hline
$\jpsi\KS$,$\psitwos\KS$,$\chicone\KS$,$\etac\KS$   & $6028$    & $92$ &  $0.713\pm0.038$   \\
$\jpsi \KL$                                         & $4323$    & $55$ &  $0.716\pm0.080$   \\
$\jpsi\Kstarz (\Kstarz \to \KS\piz)$                & $965$     & $68$ &  $0.526\pm0.284$   \\\hline
1999-2002 data                                      & $3084$    & $79$ &  $0.755\pm0.067$   \\
2003-2004 data                                      & $4850$    & $77$ &  $0.724\pm0.052$   \\
2005-2006 data                                      & $3562$    & $74$ &  $0.663\pm0.062$   \\\hline\hline
$\ \ \jpsi \KS$ ($\KS \to \pi^+ \pi^-$)             & $4076$    & $96$ &  $0.715\pm0.044$   \\
$\ \ \jpsi \KS$ ($\KS \to \pi^0 \pi^0$)             & $988$     & $88$ &  $0.581\pm0.105$   \\
$\ \ \psi(2S) \KS$ ($\KS \to \pi^+ \pi^-$)          & $622$     & $83$ &  $0.892\pm0.120$   \\
$\ \ \chicone \KS $                                 & $279$     & $89$ &  $0.709\pm0.174$   \\
$\ \ \etac\KS $                                     & $243$     & $75$ &  $0.717\pm0.229$   \\\hline
$\ $ \leptontag\ category                           & $703$     & $97$ &  $0.754\pm0.068$   \\
$\ $ \kaonitag\ category                            & $900$     & $93$ &  $0.713\pm0.066$   \\
$\ $ \kaoniitag\ category                           & $1437$    & $91$ &  $0.711\pm0.075$   \\
$\ $ \kpitag\ category                              & $1107$    & $89$ &  $0.635\pm0.117$   \\
$\ $ \piontag\ category                             & $1238$    & $91$ &  $0.587\pm0.175$   \\
$\ $ \othertag\ category                            & $823$     & $89$ &  $0.454\pm0.469$   \\\hline\hline
$B_{\rm flav}$ sample                               & $112878$  & $83$ &  $0.016\pm0.011$   \\\hline
$B^+$ sample                                        & $27775$   & $93$ &  $0.008\pm0.017$   \\\hline\hline
\end{tabular}
\end{center}
\end{table}
We determine \stwob with a simultaneous maximum likelihood fit to the 
\deltat\ distribution of the tagged $B_{\CP}$ and \bflav\ samples.
There are 65 free parameters in the fit: \stwob (1),
the average mistag fractions $\mistag$ and the
differences $\Delta\mistag$ between \Bz\ and \Bzb\ mistag fractions for each
tagging category (12), parameters for the signal \deltat\ resolution (7),
parameters for \CP\ background time dependence (8), and the difference between
$\Bz$ and $\Bzb$ reconstruction and tagging efficiencies (7); for 
\bflav\ background, time dependence (3), \deltat\ resolution
(3), and mistag fractions (24). For the \CP\ modes (except for $\jpsi \KL$), 
the apparent \CP\ asymmetry of the non-peaking background in each tagging
category is allowed to be a free parameter in the fit.  
We fix $\tau_{\Bz}=1.530\,\ps$, $\deltamd
=0.507\,\ps^{-1}$~\cite{ref:pdg2006}, $\vert \lambda \vert = 1$, and $\Delta \Gamma_d=0$.
The fit to the $B_{\CP}$ and \bflav\ samples yields
$\stwob = \fitstwob \pm \statstwob \pm \syststwob$~\cite{ref:errors}.
Figure~\ref{fig:cpdeltat} shows the \deltat\ distributions and 
asymmetries in yields between events with \Bz tags and \Bzb tags for the
$\eta_f=-1$ and $\eta_f = +1$ samples as a function of \deltat\,
overlaid with the projection of the likelihood fit result.

We perform a separate fit with only the cleanest $\eta_f=-1$ sample, in which we treat
both $\vert\lambda\vert$ and $\rm{sin}2\beta$ as free parameters. 
We do not use the modes $J/\psi K^{*0}$ and $J/\psi \KL$
to minimize the  dependence of the results on the background parametrization.
We obtain  $|\lambda| = \fitlambda \pm \statlambda \pm \systlambda$.
The updated value of $\stwob$ is consistent with the current 
world average~\cite{ref:hfag} and the theoretical estimates of the 
magnitudes of CKM matrix elements in the context of the SM~\cite{ref:CKMconstraints}. 

Ref.~\cite{ref:ciuchini} presents a model-independent study of this shift
using the measurements of \bztojpsipiz~\cite{ref:jpsipiz} to quantify the effect 
of contributions from penguin operators and long-distance contributions from penguin
contractions. They find that the deviation of the measured $S_{\CP}$ term from $\stwob$, 
$\Delta\S_{\jpsi\KS} \equiv \S_{\jpsi\KS} - \stwob\ = 0.000 \pm 0.017$
which is comparable to the systematic error from our previous publication~\cite{ref:babar2004}.
The theoretical estimates of $\Delta\S_{\jpsi\KS}$ are ${\cal O}$(10$^{-3}$)~\cite{ref:mishima}
and ${\cal O}$(10$^{-4}$)~\cite{ref:boos}.

\section{\ctwob\ from \btoccbars\ decays.}
\label{sec:ctwob}
The analysis of $\btoccbars$ decay modes imposes a constraint on $\stwob$ only, leading to a four-fold
ambiguity in the determination of $\beta$. This ambiguity can leave possible new physics undetected
even with very high precision measurements of $\stwob$. Additional constraints are obtained from the
ambiguity-free measurement of $\ctwob$ using the angular and time-dependent asymmetry in $\bpsikst$ decays
and the time-dependent Dalitz plot analyses of $\Bztodsthz$ and $\Bztodstdstks$.
The \babar\ $\Bztodsthz$ analysis is described in Refs.~\cite{ref:cheng} and~\cite{ref:gaz}. 
\subsection{$\Bztodstdstks$.}
\label{sec:ddk}
Both $\Bz$ and $\Bzb$ mesons can decay to the same final state $\Dstarp\Dstarm\KS$ via a
process dominated by the single weak phase $W$-emission $b\to\ccbar s$ transition.
Possible penguin contributions are neglected and it is therefore assumed that 
there is no direct \CP\ violation. According to Ref.~\cite{ref:browder}, 
the decay can be divided into two half Dalitz planes $s^+\le s^-$ and $s^+\ge s^-$,
where $s^+\equiv m^2(\Dstarp\KS)$ and $s^-\equiv m^2(\Dstarm\KS)$, such that the
time-dependent decay rate asymmetry of $\Bz\to\Dstarp\Dstarm\KS$ is
\begin{eqnarray*}
A(t)&\equiv&\frac{\Gamma_{\Bzb}-\Gamma_{\Bz}}{\Gamma_{\Bzb}+\Gamma_{\Bz}}
=\eta_y\frac{J_c}{J_0}\cos(\Delta m_dt)- \nonumber\\
&&\left(\frac{2J_{s1}}{J_0}\stwob +\eta_y \frac{2J_{s2}}{J_0}\ctwob\right)
\sin(\Delta m_dt),\nonumber
\end{eqnarray*}
where $\eta_y=-1(+1)$ for $s^+\le s^- (s^+\ge s^-)$.
The parameters $J_0$,$J_c$,$J_{s1}$ and $J_{s2}$ are the integrals over the half Dalitz phase space 
with $s^+<s^-$ of the functions $|a|^2+|\bar{a}|^2$, $|a|^2-|\bar{a}|^2$,
$\mathop{\cal R\mkern -2.0mu\mit e}(\bar{a}a^*)$ and
$\mathop{\cal I\mkern -2.0mu\mit m}(\bar{a}a^*)$,
where $a$ and $\bar{a}$ are the decay amplitudes of
$\Bz\to\Dstarp\Dstarm\KS$ and $\Bzb\to\Dstarp\Dstarm\KS$, respectively.
If the decay $\Bz\to\Dstarp\Dstarm\KS$ has only a non-resonant component,
the parameters $J_{s2}=0$ and $J_c$ are at the few percent level~\cite{ref:browder}.
The \CP\ asymmetry can be extracted by fitting the
\Bz\ time-dependent decay distribution. The measured \CP\ asymmetry is
\stwob multiplied by a factor of $2J_{s1}/J_0$ because the final state is an
admixture of \CP\ eigenstates with different \CP\ parities. In this
case, the value of the dilution factor $2J_{s1}/J_0$ is estimated to
be large~\cite{ref:browder}, similar to the decay $\Bz\to\Dstarp\Dstarm$.
The situation is more complicated if intermediate resonances such as
$D^+_{sJ}$ are present.  
In this case, the parameter $J_{s2}$ is non-zero and $J_c$ can be large.
The resonant components are expected to be dominated by two $P$-wave excited
$D_{s1}$ states~\cite{ref:browder}. One such state is $D^+_{s1}(2536)$ that
 has a narrow width and does not contribute much to $J_{s2}$. 
The other $D^+_{s1}$ resonant state is predicted in the quark model~\cite{ref:godfrey} 
to have a mass above the $\Dstarp\KS$ mass threshold with a large
width. In this case, the $J_{s2}$ can be large. Therefore
by studying the time-dependent asymmetry of $\Bz\to\Dstarp\Dstarm\KS$ in
two different Dalitz regions, the sign of $\cos2\beta$
can be determined for a sufficiently large data set using the method
described in Refs.~\cite{ref:browder,ref:ddkmethods}.
This would allow the resolution of the $\beta\to\pi/2-\beta$ ambiguity
despite the large theoretical uncertainty of $2J_{s2}/J_{0}$.
However, one of the expected $P$-wave $D^+_{s1}$ may be
the newly discovered $D^+_{sJ}(2317)$ or $D^+_{sJ}(2460)$.
These states are below the $\Dstarp\KS$ mass threshold, so they will not contribute to
the decay $\Bztodstdstks$. The fits to $\nbbddk$ decays yield:
${{\Jc}/{\Jzero}} = 0.76 \pm 0.18 \pm 0.07,$
${({2\Jsone}/{\Jzero})}\stwob = 0.10 \pm 0.24 \pm 0.06,$ and 
${({2\Jstwo}/{\Jzero})}\ctwob = 0.38 \pm 0.24 \pm 0.05$~\cite{ref:poireau}.
The measured value of ${\Jc}/{\Jzero}$ is significantly different from zero, which, according
to Ref.~\cite{ref:browder}, may indicate that there is a sizable broad resonant contribution to the
decay $\Bztodstdstks$ from an unknown $D^+_{s1}$ state with an unknown width. Under this assumption
then the measured value of ${2\Jstwo}/{\Jzero}$ implies that the sign of $\ctwob$ is preferred to be
positive at a 94$\%$ confidence level.
\subsection{$\Bztojpsikstar$.}
\label{sec:jpsikstar}
\babar\ has measured the sign of \ctwob\ in a time-dependent angular analysis of 104 
$\Bz\to\jpsi\Kstarz(\Kstar\to\KS\piz)$ decays in $\nbbjpsik$  
of data recorded between 1999 and 2002~\cite{ref:cos2beta}. 
Interference between decays to \CP-even (L=0,2) and \CP-odd (L=1)
final states give terms proportional to \ctwob\ in the decay rate. Strong phase differences 
and transversity amplitudes A, that appear also in these terms, have been separately measured
in a time-integrated angular analysis of $\Bpm\to\jpsi K^{*\pm}$ and $\jpsi\Kstarz(\Kstar\to K^{+}\pi^{-})$ 
decays:
\begin{eqnarray}
\delta_\parallel-\delta_0 &=& (-2.73 \pm 0.10 \pm 0.05) \rad, \nonumber \\
\delta_\perp-\delta_0 &=& (+2.96 \pm 0.07 \pm 0.05) \rad, \nonumber \\
\vert A_0 \vert^2 &=&  0.566 \pm 0.012 \pm 0.005, \nonumber \\
\vert A_\parallel \vert^2 &=&  0.204 \pm 0.015 \pm 0.005, \nonumber \\
\vert A_\perp\vert^2 &=&  0.230 \pm 0.015 \pm 0.004.\nonumber
\label{eqn:trans_final_result}
\end{eqnarray}
The analysis in principle allows a second solution for the strong phase differences, leading to a sign ambiguity in
\ctwob. This ambiguity has been resolved with the inclusion of S-wave $K\pi$ final states in the analysis. 
The interference between the S-wave and P-wave contributions gives additional terms in the decay rates
with a clear dependence on the $K\pi$ mass due to the resonance shapes. 
The other solution for the strong phase differences can be excluded as leading to an unphysical dependence 
of the strong phase differences on the $K\pi$ mass~\cite{ref:verderi}.
Using the values from Eq. \ref{eqn:trans_final_result}, and fixing \stwob to 0.731 
{\footnote{This was the world average at the time - see Ref.~\cite{ref:pdg2004}.}}, 
the fit to the $\Bz\to\jpsi\Kstarz(\Kstarz\to\KS\piz)$ sample gives
$\ctwob =+2.72_{-0.79}^{+0.50} \pm 0.27$.
By comparing this result with the outcome of fits to 2000 data-sized Monte Carlo samples,
the sign of $\cos2\beta$ is determined to be positive at the 86\% confidence level, 
in agreement with Standard Model expectations.
\section{Conclusions.}
\label{sec:conc}
When the \babar\ measurement of $\stwob$ using $\btoccbars$ decays is combined with the 
most recent Belle result described in Ref.~\cite{ref:belle} then the world average value of
$\stwob$ from $\btoccbars$ decays is $\stwob = $ 0.674 $\pm$ 0.026. 
The combined constraint on $\beta$ in the $\rhobar$-$\etabar$ plane from the \babar\ and Belle 
$\btoccbars$ charmonium + $K^{0}$ meson analyses~\cite{ref:sin2b,ref:belle}, 
the $\bpsikst$~\cite{ref:cos2beta,ref:bellejpsikstar}, $\Bztodsthz$~\cite{ref:cheng,ref:belledhzero} 
and $\Bztodstdstks$~\cite{ref:poireau} analyses strongly favour the solution 
$\beta = $21.1$\pm$1.0$^\circ$ where $\ctwob$ is positive~\cite{ref:hfag}.

\begin{acknowledgments}
The author wishes to thank the conference organizers for an enjoyable and well-organized workshop.
This work is supported by the United Kingdom Particle Physics and Astronomy Research Council (PPARC) and the 
United State Department of Energy (DOE) under contract DE-AC02-76SF00515.
\end{acknowledgments}



\begin{thebibliography}{99}
\bibitem{ref:charge}
Charge-conjugate reactions are included implicitly unless otherwise specified.
\bibitem{ref:errors}
Unless otherwise stated, all results are quoted with the first error being statistical and the second systematic.
\end{thebibliography}
\end{document}